\documentstyle[prl,aps,floats,epsf]{revtex}\def\narrowtext{}\tighten\twocolumn
\input epsf.sty
\begin{document}
\draft

\title{
$^{23}$Na NMR Studies of Non-superconducting and Superconducting $\mathrm{Na_{0.35}CoO_{2} \cdot 1.3H_{2}O}$
($T\mathrm{_{c}}<$ 1.8 and $\sim$4.5 K) }
\author{
H. Ohta, Y. Itoh, C. Michioka and K. Yoshimura
}

\address{
Department of Chemistry, Graduate School of Science,\\
Kyoto University, Kyoto 606-8502, Japan \\ 
}

\date{\today}%
\maketitle %

\begin{abstract} 
We report $^{23}$Na NMR studies of 
the polycrystalline samples  of {\it bi}layer hydrated cobalt oxides
$\mathrm{Na_{0.35}CoO_{2} \cdot 1.3H_{2}O}$ with the superconducting transition temperatures $T\mathrm{_{c}}<$1.8 and $\sim$4.5 K.
The magnitude of the $^{23}$Na nuclear spin-lattice relaxation rate 1/$^{23}T_1$ significantly decreased 
from non-hydrated Na$_{0.7}$CoO$_2$ to the {\it bi}layer hydrates. 
We found that the temperature dependence of 1/$^{23}T_1$ of the non-superconducting {\it bi}layer hydrate
is similar to that of the non-hydrated Na$_{0.7}$CoO$_2$, 
the spin dynamics of which is understood by 
$A$-type (intra-layer ferromagnetic and inter-layer antiferromagnetic) spin fluctuations.
The superconducting phase is located close to the $A$-type magnetic instability.  
\end{abstract}

\narrowtext 

The discovery of $bi$layer hydrated cobalt oxide superconductor $\mathrm{Na_{0.35}CoO_{2} \cdot 1.3H_{2}O}$ \cite{Takada}
has provided us with a new opportunity to study the itinerant electron magnetism on a triangular lattice. 
Spin frustration effect on a triangular lattice has long been expected to realize 
the resonating valence bond, so-called RVB state \cite{Anderson,Baskaran}, 
and the strong thermoelectric power of the parent Na$_x$CoO$_2$ has attracted much interests 
in the vital roles of charge fluctuations of Co$^{3+}$/Co$^{4+} $ \cite{Terasaki} 
and of the spin entropy. 
At first, the effect of water intercalation on superconductivity was thought only to extend the interlayer distance 
between the CoO$_2$ layers, 
but now it turns out that the diversity of the site occupation
of Na ions \cite{CavaNature,CavaPRB} and H$_3$O$^{+}$ oxonium ions \cite{Takada2} gives a variety of electronic states of $bi$layer hydrated cobalt oxides. 
Although the Na ions surrounded by the intercalated water molecules are 
the key ingredients for occurrence of superconductivity,  
their electronic effects on the CoO$_2$ planes are poorly understood.  

The parent non-hydrate Na$_{0.7}$CoO$_2$ has an itinerant electronic system. 
Inelastic neutron scattering studies revealed the $A$-type spin fluctuations on the CoO$_2$ planes, 
i.e. intra-plane ferromagnetic and inter-plane antiferromagnetic instability
at low temperatures \cite{Boothroyd}.
The temperature and magnetic-field dependence of electrical resistivity shows characteristic of a magnetic quantum critical point \cite{Taillefer}. 
The $^{23}$Na nuclear spin-lattice relaxation time indicates
the existence of two dimensional ferromagnetic spin fluctuations at low temperatures \cite{Ihara}
and of charge fluctuations or the motion of Na ions at high temperatures \cite{Gavilano}. 
The detail analysis of $^{23}$Na NMR spectra revealed 
the vital role of charge degree of freedom \cite{Alloul}.

A non-superconducting $bi$layer hydrate cobalt oxide, $\mathrm{Na_{0.35}CoO_{2} \cdot 1.3H_{2}O}$, 
was reported to exhibit a small internal magnetic field \cite{Ihara2}. 
The weak magnetic behavior suggests a close interplay between 
the occurrence of superconductivity and magnetic instability. 
The delicate amounts of Na ions, water molecules, and oxonium ions may control 
the carrier doping level of the CoO$_2$ planes. 
However,  
no chemical treatment to synthesize the non-superconductors has been established so far 
for the $bi$layer hydrates. 

Recently, we found successful synthesis method of the non-superconducting $bi$layer hydrate with $T\mathrm{_{c}}<$1.8 K 
and the derivative superconducting hydrate with $T\mathrm{_{c}}\sim$4.5 K with a small upper critical field $H_{c2}$. 
In this Letter, we report $^{23}$Na NMR studies for the non-superconducting 
and the derivative superconducting $bi$layer hydrates.
We observed that the Na nuclear spin-lattice relaxation rate 1/$^{23}T_1$ 
of the non-superconducting $bi$layer hydrate is of two order smaller than that of the parent Na$_{0.7}$CoO$_2$ 
but shows the similar temperature dependence. 
The electronic state of the non-superconducting $bi$layer hydrate is found to be nearly the same as that of the parent Na$_{0.7}$CoO$_2$, 
i.e. the $A$-type spin fluctuations on the CoO$_2$ planes at low temperatures
and charge fluctuations/the motion of Na ions at high temperatures. 

The parent compound, Na$_{0.7}$CoO$_2$ was synthesized by a conventional solid-state reaction. The powder of Na$_{0.7}$CoO$_2$ was immersed in Br$_2$/CH$_3$CN solution for 1 day for deintercalation of Na$^+$. The powder was immersed in distilled water for 1 day to intercalate H$_2$O molecules. After filtering the powder, the sample was put into a chamber with an atmosphere of 75 \% humidity air. Recently we have investigated the duration (keeping time in the humidity-controlled chamber) dependence of the physical properties of $bi$layer hydrates \cite{Ohta}. The sample shows a systematic change in the chamber as a function of the duration day by day. After each duration, cryopreservation at -10 $^{\circ}$C was found to be effective to quenche further change of the sample. The duration effect appears especially in the superconductivity, i.e. the superconducting transition temperature $T\mathrm{_{c}}$ and the volume fraction. In addition, the duration effect was found to be sensitive to the initial condition of the powder and the atmosphere in the chamber. 

\begin{figure}
\epsfxsize=2.8in
\epsfbox{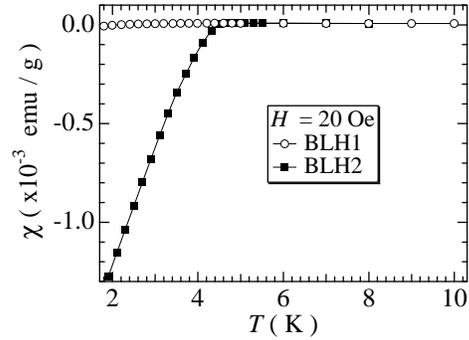}
\vspace{0.0cm}
\caption{\label{fig:chi}
Temperature dependence of $\chi$ of the $bi$layer hydrated BLH1 and BLH2 after cooling in zero magnetic field. $T\mathrm{_{c}}$ of BLH2 is $\sim 4.5$ K, but that of BLH1 is $< 1.8$ K.
}
\end{figure}

 In this study we prepared two samples. The duration of the sample named BLH1 is 1 week, and that of BLH2 is 1 month. 
Figure \ref{fig:chi} shows the temperature dependence of the magnetic susceptibility under $H$ = 20 Oe measured by a SQUID magnetometer. $T\mathrm{_{c}}$ of BLH2 is 4.5 K while no signs of the superconductivity appear down to 1.8 K in BLH1.
Any samples with short duration do not show superconductivity down to 1.8 K, and then BLH1 is classified into these non-superconducting $bi$layer hydrates.
The present BLH2 has a smaller upper critical field $H_{\mathrm{c2}}$ than that reported in Ref. \cite{Sakurai}.
Zero-field $^{59}$Co nuclear quadrupole resonance (NQR) experiments have also been performed to characterize the samples. For BLH1, the appearance of an internal magnetic field below about 4.5 K was observed in a broad $^{59}$Co NQR spectrum, being similar to the previous report \cite{Ihara2}. For BLH2, the superconducting transition was confirmed by the rapid decrease of the $^{59}$Co nuclear spin-lattice relaxation rate 1/$^{59}T_1$. The details of the results will be reported elsewhere.

The $^{23}$Na (spin $I$=3/2) NMR experiments have been carried out 
by a coherent-type pulsed NMR spectrometer for the powdered samples.      
The $^{23}$Na NMR frequency spectra 
were obtained by a fast Fourier transformation (FFT) technique 
for the free induction signals.
The $^{23}$Na nuclear spin-lattice relaxation time $^{23}T_{1}$ was measured 
by an inversion recovery technique. 
The free induction signal $M(t)$ as a function of a duration time $t$ after an inversion pulse 
and $M(\infty)[\equiv M(t>10T_1)]$ were recorded. 
In order to estimate $^{23}T_{1}$, 
the recovery curves
$p(t)\equiv 1-M(t)/M(\infty)$ were fit to the data by
$p(t)=p(0)[0.1\mathrm{e}^{-6t/{T_1}}+0.9\mathrm{e}^{-3t/{T_1}}]$
for central transition ($I_z$=1/2$\leftrightarrow$-1/2),
where $p(0)$ and $T_1$ are fit parameters.   

\begin{figure}
\epsfxsize=2.8in
\epsfbox{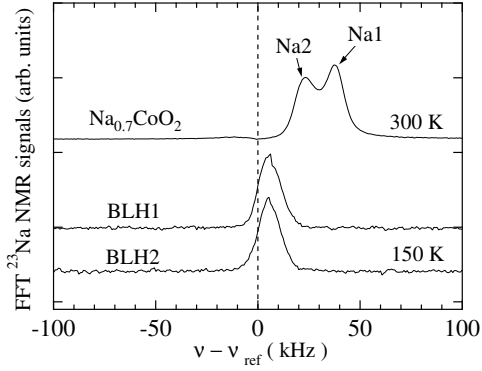}
\vspace{0.0cm}
\caption{\label{fig:FFT}
Fast Fourier transformed $^{23}$Na NMR frequency spectra of a parent Na$_{0.7}$CoO$_{2}$,
$bi$layer hydrated BLH1 and BLH2.  
The zero shift indicated by the dashed line is
$\nu_{\mathrm{ref}}$=75.2542 MHz 
($\nu_{\mathrm{ref}}$=$^{23}\gamma_{n}H$ 
with $^{23}\gamma_{n}$=10.054 MHz/T and $H\sim$7.485 T),
referred to the $^{23}$Na NMR line of NaCl in water at 300 K.    
}
\end{figure}

Figure \ref{fig:FFT} shows the FFT $^{23}$Na NMR spectra of the parent Na$_{0.7}$CoO$_{2}$, the $bi$layer hydrated BLH1 and BLH2.
For Na$_x$CoO$_2$, there are two crystallographic Na sites, Na1 and Na2, 
which reside just above the Co site and in the center of the Co triangles, respectively.       
The two $^{23}$Na NMR lines in Na$_x$CoO$_2$ in Fig. \ref{fig:FFT} are assigned to the central transition lines ($I_z$=1/2$\leftrightarrow$-1/2) of Na1 and Na2 nuclei \cite{Alloul}.
They are affected by finite Knight shift and electric quadruople shift, which is consistent with Ref.\cite{Alloul}. 
For the $bi$layer hydrates, the single sharp lines with small Knight shift are observed, being similar to the result for Na$_{0.35}$CoO$_2$ (sodium deintercalated non-hydrate compound) in Ref. \cite{Alloul}.  
The hydration diminishes the Co-to-Na hyperfine coupling and shields the electric field gradient from
the charge distribution on the CoO$_2$ planes. 
The $^{23}$Na NMR spectra and nuclear spin-lattice relaxation studies (shown below) 
revealed the electric and magnetic shielding effect by water intercalation. 

\begin{figure}
\epsfxsize=2.8in
\epsfbox{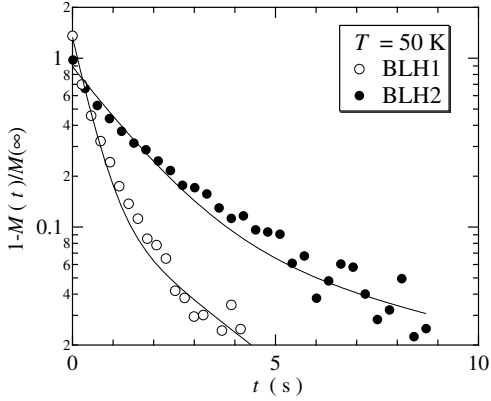}
\vspace{0.0cm}
\caption{\label{fig:rec}
$^{23}$Na nuclear spin-lattice relaxation curves (recovery curves) 
for $bi$layer hydrated BLH1 and BLH2 at 50 K.
The solid curves are the least-squares fits by theoretical curves (see text)
for central transition lines ($I_z$=1/2$\leftrightarrow$-1/2).    
}
\end{figure}

Figure \ref{fig:rec} shows typical $^{23}$Na nuclear spin-lattice relaxation curves (recovery curves) for $bi$layer hydrated BLH1 and BLH2 at 50 K. 
The solid curves are the least-squares fits by the theoretical function for central transition lines ($I_z$=1/2$\leftrightarrow$-1/2) mentioned above. 
Non-single exponential relaxation was observed in all the recovery curves. 
Since neither appreciable split nor asymmetric profile due to the electric quadrupole shift was observed in the present $^{23}$Na NMR spectra in Fig. \ref{fig:FFT}, 
then we assumed that the satellite lines of $I_z$=$\pm$3/2$\leftrightarrow$$\pm$1/2 are wiped out.
Thus, we determined the values of $^{23}T_1$ from these fits. 
In passing, we have already analyzed the data by a stretched exponential function based on an impurity-induced NMR relaxation theory \cite{Mac}.  
We confirmed that the present results on $T_1$ are independent of the details of analysis. 

\begin{figure}
\epsfxsize=3.5in
\epsfbox{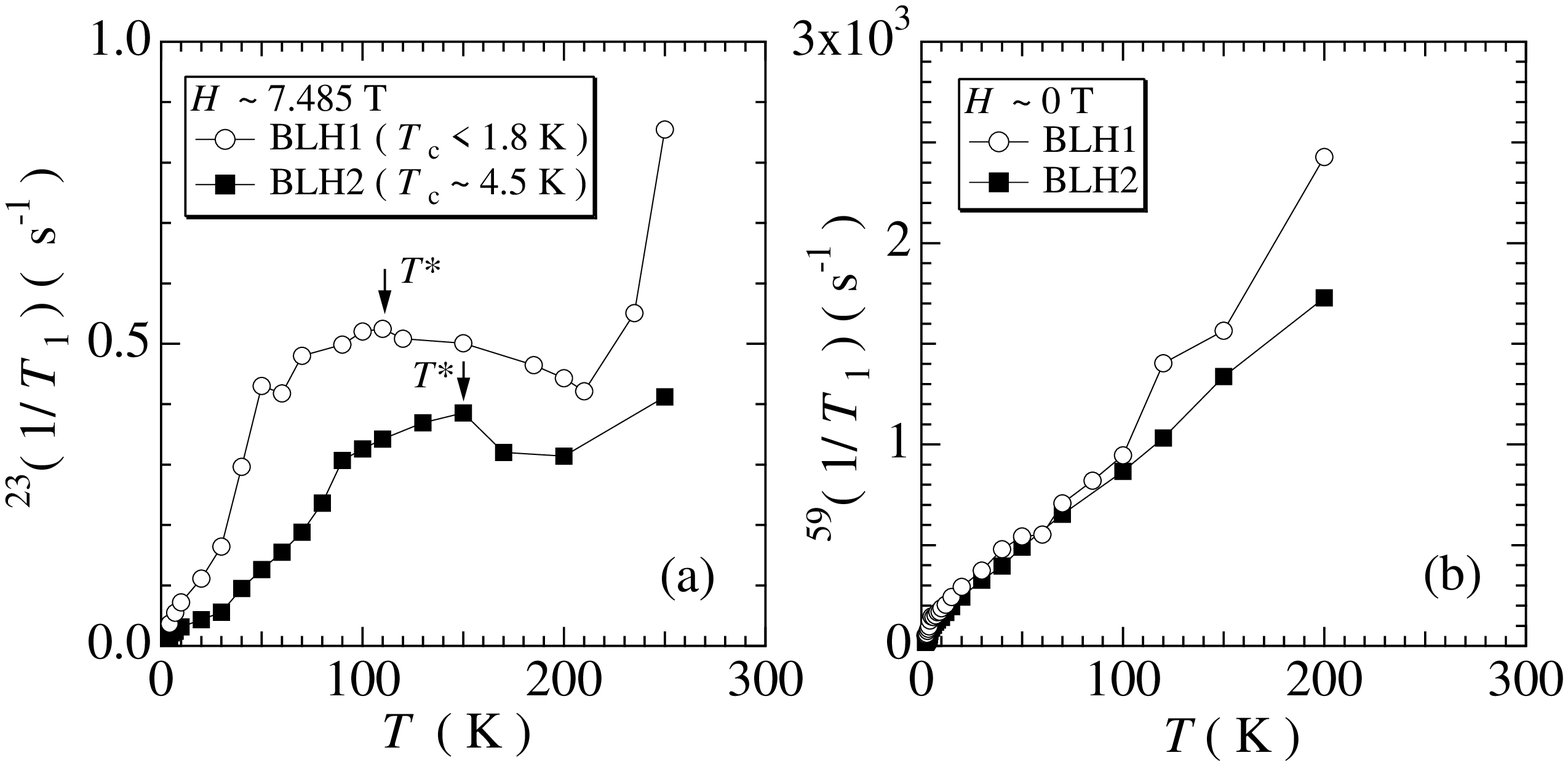}
\vspace{0.0cm}
\caption{\label{fig:invT1}
(a) Temperature dependence of 
the $^{23}$Na nuclear spin-lattice relaxation rate 1/$^{23}T_{1}$ 
for BLH1 and BLH2. 
(b)  Temperature dependence of 
the $^{59}$Co nuclear spin-lattice relaxation rate 1/$^{59}T_{1}$ 
for BLH1 and BLH2. 
} 
\end{figure}

Figure \ref{fig:invT1}(a) shows the temperature dependence of the $^{23}$Na nuclear spin-lattice relaxation rate 1/$^{23}T_{1}$ for BLH1 and BLH2.  
For comparison, the temperature dependence of 
the $^{59}$Co nuclear spin-lattice relaxation rate 1/$^{59}T_{1}$ for BLH1 and BLH2 are shown in Fig. \ref{fig:invT1}(b).
Since all the 1/$^{23}T_{1}$ seem to reach zero at $T$=0 K in Fig. \ref{fig:invT1}(a), no paramagnetic impurity causes the $^{23}$Na NMR relaxation. 
The behavior of 1/$^{23}T_{1}$ in Fig. \ref{fig:invT1}(a) can considered to be intrinsic. 
The magnitude of 1/$^{23}T_{1}$ significantly decreases from BLH1 ($T\mathrm{_{c}}<$1.8 K) to BLH2 ($T\mathrm{_{c}}$=4.5K),
whereas that of 1/$^{59}T_{1}$ at low temperatures does not change so much. 
The broad maxima of 1/$^{23}T_{1}$, denoted by $T^{*}$, are seen in Fig. \ref{fig:invT1}(a),
whereas 1/$^{59}T_{1}$ is not in Fig. \ref{fig:invT1}(b).  
We suppose that the different $T\mathrm{_{c}}$ results from the difference in the degree of hydration.
Thus, the difference in spin dynamics at the Na site indicates the difference in shielding effect. 
The different temperature dependence of 1/$^{23}T_{1}$ and 1/$^{59}T_{1}$ indicates the existence of an another fluctuation mode at the Na site between the CoO$_2$ planes,
in addition to transferred fluctuations from the in-plane electron spin fluctuations.  
The in-plane spin fluctuations probed by 1/$^{59}T_{1}$ is nearly independent of $T\mathrm{_{c}}$'s,
whereas an interplane fluctuations probed by 1/$^{23}T_{1}$ is strongly dependent on $T\mathrm{_{c}}$'s.
Thus, an interlayer coupling probed by the Na NMR $^{23}T_1$ may control the value of $T\mathrm{_{c}}$. 

For the non-superconducting BLH1, the increase of 1/$^{23}T_{1}$ at high temperatures is similar to that for the non-hydrate Na$_{0.7}$CoO$_2$ \cite{Gavilano}. 
After Gavilano $et$ $al.$ \cite{Gavilano}, the rapid increase of 1/$^{23}T_{1}$ above 200 K can be attributed to the motion of the Na ions. 
The power-law decrease $T^{0.7}$
of 1/$^{23}T_{1}$ at low temperatures (shown below) agrees with 
that for the non-hydrate Na$_{0.7}$CoO$_2$ \cite{Ihara}.
 
\begin{figure}
\epsfxsize=2.8in
\epsfbox{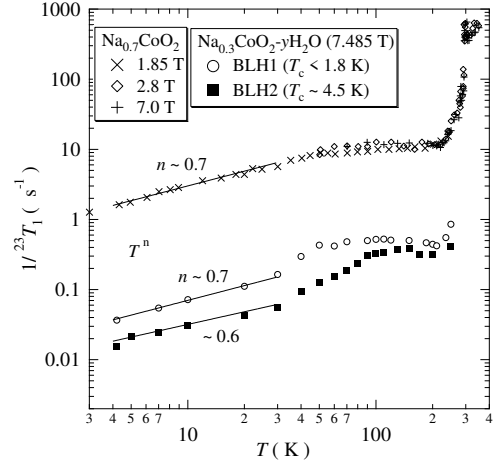}
\vspace{0.0cm}
\caption{\label{fig:Log}
Log-log plots of 
$^{23}$Na nuclear spin-lattice rate 1/$^{23}T_1$ against temperature
for the non-hydareted Na$_{0.7}$CoO$_2$, which are reproduced from
Ref. [10, 11], the $bi$layer hydrated $\mathrm{Na_{0.35}CoO_{2} \cdot 1.3H_{2}O}$, BLH1 and BLH2. 
The solid lines are fits to the low temperature 1/$^{23}T_1$ by the power laws of $T^{n}$.
}
\end{figure}

Figure \ref{fig:Log} shows the log-log plots of 1/$^{23}T_1$ against temperature for the non-hydrated Na$_{0.7}$CoO$_2$, which are reproduced from Ref.\cite{Ihara,Gavilano}, and for the $bi$layer hydrated BLH1 and BLH2. 
In all the plots, non-Korringa law is observed. 
As indicated by the solid lines in Fig. \ref{fig:Log}, all the 1/$^{23}T_1$'s at low temperatures show the power-law behaviors of $T^{n}$ with $n\sim$0.7 and 0.6 for BLH1 and BLH2, respectively. 
The magnitude of 1/$^{23}T_{1}$ significantly decreases from the non-hydrated to the $bi$layer hydrated samples. 
The water intercalation causes the shielding effect on the Co-to-Na transferred hyperfine coupling as mentioned above.

\begin{figure}
\epsfxsize=3.0in
\epsfbox{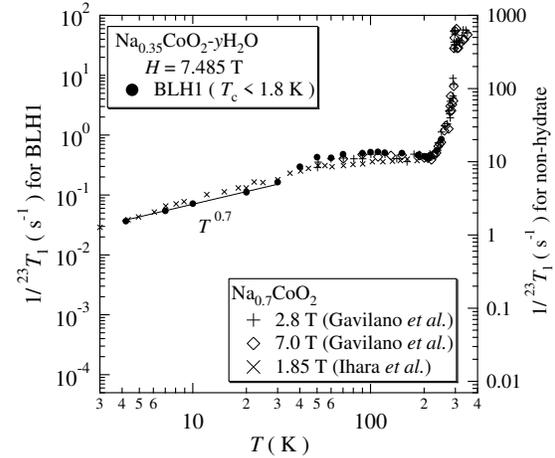}
\vspace{0.0cm}
\caption{\label{fig:scaling}
Temperature dependence of 
the $^{23}$Na nuclear spin-lattice relaxation rate 1/$^{23}T_{1}$ 
for the parent Na$_{0.7}$CoO$_2$ and the $bi$layer hydrated non-superconductor BLH1. 
}
\end{figure}

Figure \ref{fig:scaling} shows the log-log plot of 1/$^{23}T_{1}$ against temperature for the parent Na$_{0.7}$CoO$_2$ and the non-superconductor BLH1.
The magnitude is different with each other but the temperature dependence is nearly the same in a whole temperature region. 
The ratio of 1/$^{23}T_{1}$ of $\mathrm{Na_{0.7}CoO_{2}}$ to that of BLH1 is $\sim$45 below 10 K. In general, 1/$^{23}T_{1}$ is expressed by $\propto H\mathrm{_{hf}}^{2}S(\omega_{n})$ with the hyperfine coupling constant $H\mathrm{_{hf}}$ and a spin-fluctuation spectrum $S(\omega_{n})$ at an NMR frequency 2$\pi \omega_{n}$ \cite{Moriya}. The difference in the magnitude of 1/$^{23}T_{1}$ can be ascribed to that in $H\mathrm{_{hf}}$, whose ratio is about 6.6 below 10 K.
Thus, the spin dynamics, $S(\omega_{n})$, at the Na site of the non-superconducting BLH1 is similar to that of the non-hydrated Na$_{0.7}$CoO$_2$. 
The $A$-type spin fluctuations, i.e. intra-plane ferromagnetic and inter-plane antiferromagnetic fluctuations, are observed in the non-hydrated Na$_{0.7}$CoO$_2$ \cite{Boothroyd}. 
The intercalated water molecules may cut the inter-plane antiferromagnetic couplings.
Nearly ferromagnetic intra-plane spin fluctuations persist in the $bi$layer hydrates. 

In conclusion, the intercalated water molecules act as the shielding effect on the Co-to-Na magnetic
and electric coupling. 
The spin fluctuations of the $bi$layer hydrated non-superconductor are similar to
the $A$-type spin fluctuations of the non-hydrated Na$_{0.7}$CoO$_2$.  
The superconductivity appears close to the $A$-type magnetic instability. 

\acknowledgments
We thank Mr. T. Waki and Dr. M. Kato for their experimental supports,
and Dr. K. Ishida for fruitful discussions.
This paper is supported in part 
by Grants-in Aid for Scientific Research
from Japan Society for the Promotion of
Science (16076210).

\end{document}